\documentclass[aps,pre,amsmath,amssymb,showpacs,amsfonts]{revtex4}
\usepackage{epsfig}
\usepackage{graphicx}
\usepackage{dcolumn}
\usepackage{bm}
\usepackage{amsmath, amsthm, amssymb}
\usepackage{subfigure}
\usepackage{color}

\newtheorem{statement}{Statement}
\newtheorem*{reformulation}{Reformulation}
\newtheorem*{conjecture}{Conjecture}

\begin{document}

\title{Codimension-2 surfaces and their Hilbert spaces: low-energy clues for holography from general covariance}
\author{Yakov Neiman}
\email{yashula@gmail.com}
\affiliation{Raymond and Beverly Sackler School of Physics and Astronomy, Tel-Aviv University, Tel-Aviv 69978, Israel}
\eprint{0901.4048}

\date{\today}

\begin{abstract}
We argue that the holographic principle may be hinted at already from low-energy considerations, assuming diffeomorphism invariance, quantum mechanics and Minkowski-like causality. We consider the states of finite spacelike hypersurfaces in a diffeomorphism-invariant QFT. A low-energy regularization is assumed. We note a natural dependence of the Hilbert space on a codimension-2 boundary surface. The Hilbert product is defined dynamically, in terms of transition amplitudes which are described by a path integral. We show that a canonical basis is incompatible with these assumptions, which opens the possibility for a smaller Hilbert-space dimension than canonically expected. We argue further that this dimension may decrease with surface area at constant volume, hinting at holographic area-proportionality. We draw comparisons with other approaches and setups, and propose an interpretation for the non-holographic space of graviton states at asymptotically-Minkowski null infinity. 
\end{abstract}

 
\maketitle

\section{Motivation}

It seems difficult to formulate a theory that respects both quantum mechanics and diffeomorphism invariance, while allowing to describe processes which take place over bounded regions of space and time. Existing attempts typically either describe observables only at the spacetime's boundary, or fail to preserve diffeomorphism invariance at the quantum level. Several of these will be addressed below. It may be that diffeomorphism invariance is merely an emergent symmetry, making an invariant quantum theory unnecessary. Or it may be that such a theory exists, if only as a low-energy limit, and is yet to be discovered. Assuming that this is the case, we can try to extract some of the theory's general features from its desired properties. This is the approach taken in the present paper. We propose an apparently natural description of low-energy states and observables in the theory. As a purpose and a guiding line for the discussion, we present some arguments, increasingly specific yet increasingly speculative, implying that this description is consistent with the gravitational holographic principle. If true, this suggests that the holographic principle follows from quantum mechanics together with diffeomorphism invariance, not requiring the specific form of Einsteinian gravity.

The holographic principle states that the number of degrees of freedom of a quantum gravitational system in a volume $V$ is proportional to its surface area $A$, with one degree of freedom per Planck cell; this stands in contrast to the naive picture of QFT with a cutoff at Planck mass $M_P$, which would suggest a volume-proportionality. The principle has been first proposed based on thought experiments regarding black holes and saturation of the entropy bound, within the framework of conventional general relativity and particle theory \cite{Bekenstein:2000ai,'tHooft:1999bw,Susskind:1994vu}. Initially non-covariant, the principle has subsequently been framed in several covariant versions and formulations; see \cite{Bousso:1999dw,Flanagan:1999jp,Brustein:1999md}. These have also been tested by thought experiments in classical General Relativity.

Microscopic evidence in favor of the principle has been accumulated via the gauge/gravity (string) duality that relates a quantum gravitational (string) theory on a $d$-dimensional asymptotically Anti de-Sitter (AdS) space to a quantum gauge (field) theory without gravity on its $(d-1)$-dimensional boundary \cite{Aharony:1999ti}. Of particular importance there are the examples of charged black holes, whose microscopic description in terms of D-branes allowed a counting of their degrees of freedom in agreement with the holographic Bekenstein-Hawking formula. In loop quantum gravity, a successful state-counting with the same result has been achieved for more general black holes; see \cite{Rovelli:1996ti} for a review.

These pieces of evidence are often disjoint, and at times esoteric. One is tempted to look for simple, general arguments that don't go too far into new theory. We could attempt such an argument involving nothing but a scalar QFT in a box. Apart from the microscopic Planck cutoff, we may introduce a macroscopic bound: the energy in the box should be below the mass of a black hole with the box's size. Assuming a thermal state with the appropriate energy $E$, we arive at an entropy $\propto \sqrt{V}$, which for spacetime dimension $d \geq 4$ is even lower than the holographic value. Note that for a large box, we have $T \ll M_P \ll E$, where $T$ is the saturation temperature. Thus the Planck cutoff isn't even approached, since the (much lower) temperature serves as the effective cutoff. A curious calculation in the same setting by Yurtsever \cite{Yurtsever:2003ii} arrives at an area-proportional entropy, by using the same Planck cutoff and energy bound, but allowing states far from thermal equilibrium.

Another line of research has been to formulate and explain the holographic principle in general diffeomorphism-invariant classical theories \cite{Wald:1993nt,Iyer:1994ys}. There, black hole entropy is associated with the Noether currents of diffeomorphisms. This suggests that the principle is related mainly to the diffeomorphism-invariant nature of gravity. 

We wish to pursue this concept of the central role played by diffeo-invariance, in a low-energy quantum-mechanical setting. We will consider a general diffeomorphism-invariant quantum field theory in finite spatial regions of spacetime with a Minkowski-like causal structure. While not requiring the exact variables and shape of General Relativity, we assume the fields to provide some measure of scale, like the metric defines length. We will be working in the weak-field regime, with quantum states describable in terms of classical geometries, such that any small enough region of spacetime has a similar topological and causal structure to a region of Minkowski space. In particular, we will not address black holes. It is interesting if holographic features appear already from these simple working assumptions.

\section{Overview}

As our primary observable, we take the transition amplitudes between field configurations on finite spacelike hypersurfaces. We require a Hilbert-space structure that is consistent with these amplitudes. In the spirit of diffeomorphism invariance, we treat the boundary conditions $B$ for an amplitude $T[B]$ as intrinsic to the boundary hypersurace, without reference to its embedding. This point of view leads to a certain 'floppiness' regarding the relative position of the past and future boundaries, which plays a central role in the argument. To express the amplitude $T[B]$, we use a path integral over the region bounded by $B$. 

By sticking to low-energy arguments, we assume that the path integral is regularized at an energy scale $\Lambda \ll M_P$ far from the Planck scale, in such a way that diffeomorphism invariance (at the appropriate scale, at least) is preserved. As already stated, for the general amplitudes we consider, this assumption is merely an educated guess, not supported by a known calculation method. But as we shall see, it allows us to formulate features of the theory that wouldn't be seen if diffeomorphism invariance wasn't assumed (Sec. \ref{sec:QFT}), or if we considered special amplitudes that are easier to regularize (Sec. \ref{subsec:observables:special}). However, such apecial amplitudes may be useful for a quantitative test of our arguments (Sec. \ref{sec:quantitative}). The low-energy regularization means that the true entropy bound with its Planck-scale coefficient isn't saturated. Instead, we look in the smaller regularized Hilbert space for features that are characteristic of the holographic bound.

A distinctive feature of our method is the use of a 'dynamical' definition for the Hilbert space: the Hilbert product is to be derived as a limit of general transition amplitudes. This is contrasted (Sec. \ref{subsec:methods:canonical}) with the 'a-priori' definition, which starts from a canonical orthogonal basis and then proceeds to derive the dynamics. For ordinary QFT (Sec. \ref{sec:QFT}), the two approaches are equivalent. However, we shall argue that their results become distinct under diffeo-invariance, leading perhaps to Hilbert spaces of different size, and that the 'dynamical' approach is more appropriate.

Within this framework, we formulate the following statements:
\begin{enumerate}
\item A codimension-2 surface $S$ is associated with a Hilbert space $H[S]$. In particular, $\dim{H}$ is a functional of the surface's properties.
\item The canonical volume-proportional basis for $H[S]$ is inconsistent with diffeomorphism-invariant transition amplitudes. This implies the possibility of a smaller basis. 
\item When the surface area is reduced without affecting the enclosed volume, $\dim{H}$ may become smaller. This is a step towards proportionality between $\dim{H[S]}$ and the surface area $A[S]$.
\end{enumerate}
The first statement follows directly from our definitions of states and observables. The second and third are more speculative.

Sec. \ref{sec:main} contains the necessary definitions, and outlines the main argument. The subsequent sections substantiate the case through comparison. Sec. \ref{sec:QFT} deals with ordinary QFT and gauge theory, showing how a parallel argument doesn't lead to holography in that case.
Sections \ref{sec:methods}, \ref{sec:observables} seek to justify our approach by considering alternatives. Sec. \ref{sec:methods} addresses some approaches to quantum gravity which appear to contradict our own: global path integrals in \ref{subsec:methods:global_local}, and canonical quantum gravity in \ref{subsec:methods:canonical}; the contradictions are shown to arise from underlying gauge-dependence of those approaches. In Sec. \ref{sec:observables} we consider other choices of observables. Our particular choice combines diffeomorphism invariance, which is absent in local field correlations (Sec. \ref{subsec:observables:correlations}), with a greater richness of structure than asymptotic correlations at infinity (Sec. \ref{subsec:observables:special}). This additional structure is shown to be essential to our argument; in particular, in Sec. \ref{subsec:observables:special} we offer some insight on the apparently non-holographic nature of the graviton Hilbert space and S-matrix in asymptotically-flat spacetime. In Sec. \ref{subsec:observables:different_causal}, we note the crucial influence of the boundary's causal structure on the theory's observables. Sec. \ref{sec:regularization} contains a brief discussion of possible mechanisms for the low-energy regularization we're assuming. In Sec. \ref{sec:quantitative}, we offer a direction for advancing our ideas to a quantitative level. 

\section{The main argument} \label{sec:main}

\subsection{Surface-associated states and transitions} \label{subsec:main:states_transitions}

Our primary object of interest is a codimension-2 spacelike surface $S$ of topology $S_{d-2}$. The definition of $S$ may include various intrinsic properties, such as its area $A[S]$ or its full intrinsic metric $g_{ab}(S)$. For scaling to make sense, we assume that we are given at least $A[S]$ as a measure of the surface's size. 

Associated with $S$ is the set $\mathbf{\Sigma}[S]$ of spacelike hypersurfaces $\Sigma$ with topology $B_{d-1}$, such that $\partial\Sigma = S$. These hypersurfaces are defined by the values taken, up to diffeomorphisms and gauge transformations, by a set of fields $\phi^{(d-1)}$: 
\begin{eqnarray}
\Sigma:\ \phi^{(d-1)}(x^i) = \phi^{(d-1)}_\Sigma(x^i);\ x^i \in B_{d-1}
\end{eqnarray}
(the superscript $(d-1)$ indicates that these fields live on a $(d-1)$-dimensional hypersurface). The field values $\phi^{(d-1)}_\Sigma(x^i)$ at $x^i \in S$ should be consistent with the fixed intrinsic properties of $S$. 

\begin{figure}%
\centering%
\label{fig:S_V_B_Omega}%
\subfigure[{The surface $S$, reduced from $d-2$ dimensions to 1.}]%
{\includegraphics{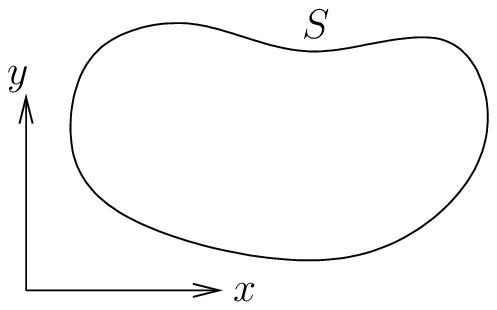}}\qquad
\subfigure[{A possible configuration $\Sigma \in \mathbf{\Sigma}[S]$, reduced from $d-1$ dimensions to 2.}]%
{\includegraphics{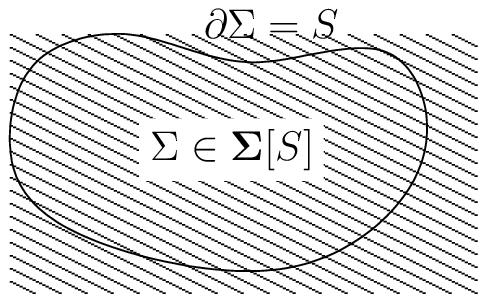}}\\
\subfigure[{A possible boundary condition $B[\Sigma_i,\Sigma_f,j]$ for the transition functional $T_S$, reduced from $d-1$ dimensions to 1; $S$ has been reduced to two points, but is actually connected.}]%
{\includegraphics{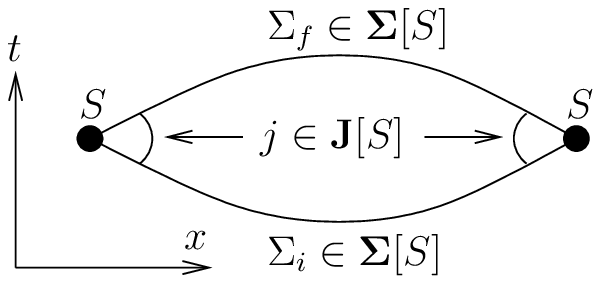}}\quad
\subfigure[{A possible path $\Omega$ that contributes to the path integral $T_S[B]$, reduced from $d$ dimensions to 2.}]%
{\includegraphics{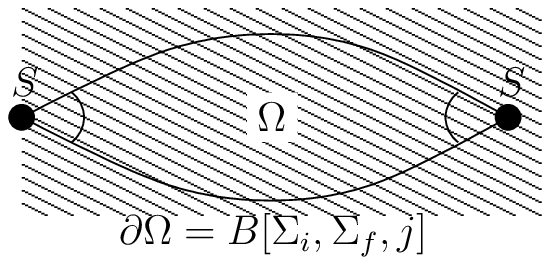}}%
\caption{Cross-section visualizations of the concepts described in Sec. \ref{subsec:main:states_transitions}. The axes indicate spacelike vs. timelike directions, and relate the cross-sections to each other; no special coordinate system is implied.}%
\end{figure}

Using appropriate fields $\phi^{(d-1)}$, we can identify the hypersurfaces $\Sigma \in \mathbf{\Sigma}[S]$ as \emph{classical configurations} of a region in space; for example, in pure gravity, the intrinsic metric $g_{ij}(\Sigma)$ would make a suitable choice of $\phi^{(d-1)}$. Each $\Sigma$ can be viewed as a restriction of $d$-dimensional fields to a Cauchy hypersurface (which defines also its geometric shape), except there is no a-priori embedding spacetime region. Our use of the condition $\partial\Sigma = S$ to define a configuration space makes sense, as it allows \emph{transitions} between such configurations. The probability amplitudes of such transitions are the observable of interest for this article. The transition amplitudes are formally described by a functional:
\begin{eqnarray}
T_S:\ \mathbf{\Sigma}[S]^2 \times \mathbf{J}[S] \rightarrow \mathbf{C} 
\end{eqnarray}
where $\mathbf{J}[S]$ is a space of 'joining parameters' $j^{(d-2)}(x^a)$; these are functions of $x^a \in S$ which describe the intersection of $\Sigma_i,\Sigma_f \in \mathbf{\Sigma}[S]$ at their mutual boundary $\partial\Sigma_i = \partial\Sigma_f = S$: for instance, $j^{(d-2)}(x^a)$ may specify the angle between $\Sigma_i$ and $\Sigma_f$ at $x^a$. Together, an initial configuration $\Sigma_i$, a final configuration $\Sigma_f$ and a joining function $j$ define a compact codimension-1 boundary $B[\Sigma_i,\Sigma_f,j]$. Like $S$ itself, this $B$ is defined intrinsically, without reference to a background spacetime \footnote{This doesn't prohibit the use of quantities like the extrinsic curvature; though related to the embedding of $B$, it's expressible within $B$'s intrinsic tensor bundle.}. Within this boundary, a path integral can be performed, yielding the transition amplitude:
\begin{eqnarray} \label{eq:transition}
T_S[B] \equiv \langle\Sigma_f|\Sigma_i\rangle_j = 
\left(\int{\frac{\mathcal{D}\phi^{(d)}_\Omega(x^\mu)}{\mbox{Diff}[\phi^{(d)}_\Omega(x^\mu)]} e^{iS[\phi^{(d)}_\Omega]}} 
\right)_{\partial\Omega = B;\ \phi_\Omega(B) = \phi_B(B)}
\end{eqnarray}
where $\phi_B$ collectively denotes the boundary conditions on the bulk fields $\phi^{(d)}_\Omega$. Fig. \ref{fig:S_V_B_Omega} shows the interrelations between the various objects in eq. \eqref{eq:transition}. We summarize our first result:
\begin{statement}
A surface $S$ is naturally associated with a configuration space $\mathbf{\Sigma}[S]$, along with a transition-amplitude functional $T_S$.
\end{statement}

The crucial role of compact hypersurfaces in defining observables for quantum gravity is discussed more generally in \cite{Oeckl:2003pw,Oeckl:2003vu}. See also \cite{Pfeiffer:2003tx} for a topological point of view. A discussion and concrete application of an expression similar to \eqref{eq:transition} can be found in \cite{Modesto:2005sj,Rovelli:2005yj}, in the context of propagators for loop quantum gravity. 

As a space of $(d-1)$-argument functions, the configuration space $\mathbf{\Sigma}[S]$ appears non-holographic, and we will treat it as such. Note that although the $\mathbf{\Sigma}[S]$ is formally continuous, it is effectively countable, due to regularization and the restriction to normalizable states of finite energy. Space is made effectively discrete at 'lattice' separation $1/\Lambda$; the field space can then be discretized by energy level, and made finite by an energy cutoff. The configuration space's effective size is then expected to scale as:
\begin{eqnarray} \label{eq:configuration_space_size}
|\mathbf{\Sigma}[S]| \sim N_\phi(\Lambda)^{\Lambda^{d-1}V[S]}
\end{eqnarray}
Here, $N_\phi(\Lambda)$ is the effective discrete volume of field space, $V[S]$ is an estimate of the $(d-1)$-dimensional volume bounded by $S$, and $\Lambda$ is the regularization scale. For a discussion on the validity of \eqref{eq:configuration_space_size}, on the difficulty in defining $V[S]$ and on the meaning of a cutoff when $\Sigma$ has no presupposed metric, see Sec. \ref{sec:regularization}. 

\subsection{Inadequacy of the canonical basis} \label{subsec:main:canonical}

Quantum mechanics suggests that the configuration space $\mathbf{\Sigma}[S]$ should span a Hilbert space $H[S]$. This involves more than the construction of linear superpositions. What determines the size and structure of the physical Hilbert space is its inner product $P\,$; it can render some superpositions of $\Sigma$'s unphysical, and others - equivalent. To deduce the inner-product structure of $H[S]$, it suffices to define it on the spanning set $\mathbf{\Sigma}[S]$:
\begin{eqnarray}
P_S:\ \mathbf{\Sigma}[S]^2 \rightarrow \mathbf{C}
\end{eqnarray}
A natural mapping $P_S \mapsto T_S$ should exist between the Hilbert inner product and the transition amplitudes: 
\begin{eqnarray}
P_S[\Sigma_1,\Sigma_2] = T_S[\Sigma_1,\Sigma_2,j_0]
\end{eqnarray}
where $j_0$ is an 'identity' joining function, such as a zero angle. This expresses the 'dynamical' definition of the Hilbert space mentioned in the Introduction. Now, the requirement of such a mapping leads to the conclusion that $\mathbf{\Sigma}$ cannot constitute an orthogonal basis for $H$. Furthermore, the same is true for a subspace $\mathbf{\Sigma}_C \subset \mathbf{\Sigma}$ subject to a Hamiltonian constraint. This is because in a $T_S$-transition, the hypersurfaces $\Sigma_i,\Sigma_f$ cannot be made a-priori to coincide: their mutual position is constrained only by the surface $S$ and the joining parameters $j$. Therefore, for any pair $\Sigma_i,\Sigma_f$, there exist finite paths $\phi^{(d)}_\Omega(x^\mu)$ between the two states; this leads in general to a nonzero value for $\langle\Sigma_f|\Sigma_i\rangle_j$. As a result, the space $\mathbf{\Sigma}$ of magnitude $N_\phi(\Lambda)^{\Lambda^{d-1}V[S]}$, or $\mathbf{\Sigma}_C$ of magnitude $N_{\phi,C}(\Lambda)^{\Lambda^{d-1}V[S]}$,
need not be a basis for $H[S]$. Since $\mathbf{\Sigma}$ is still a spanning set, this suggests the possibility of a \emph{smaller} basis, perhaps one with the holographic magnitude $\tilde{N}_\phi(\Lambda)^{\Lambda^{d-2}A[S]}$. Here, $N_{\phi,C}(\Lambda)$ and $\tilde{N}_\phi(\Lambda)$ are appropriate field-space volumes. To summarize our second result:
\begin{statement}
The canonical configuration space $\mathbf{\Sigma}_C$ is not an orthogonal basis for $H[S]$. This legitimizes a smaller, perhaps holographic, value for $\dim{H[S]}$.
\end{statement}

We stress that the Hilbert dimension discussed here is in any case smaller than the true entropy bound, due to the low-energy regularization.

\subsection{Area-dependence of the Hilbert dimension} \label{subsec:main:proportionality}

\begin{figure}%
\centering%
\subfigure[{A surface $S_{123}$, composed of two overlapping compact pieces $S_{12}$, $S_{23}$.}] 
{\label{subfig:S_123} \includegraphics{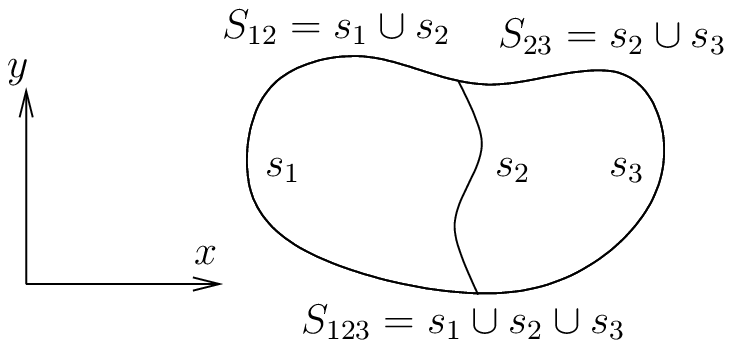}} \qquad
\subfigure[{The surface $S_{13}$ obtained by removing the overlap segment $s_2$}]
{\label{subfig:S_13} \includegraphics{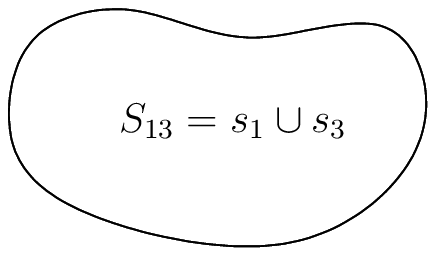}} \\
\subfigure[{A possible path $\Omega$ between a pair of configurations $\Sigma_i,\Sigma_f \in \mathbf{\Sigma}[S_{123}]$. This path contributes to $S_{123}$- as well as $S_{13}$-associated transitions.}]
{\label{subfig:B_123} \includegraphics{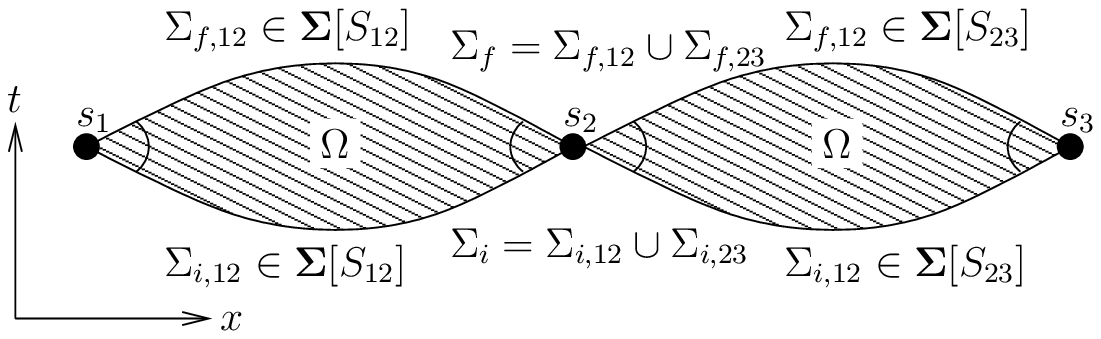}} \\
\subfigure[{A different path $\Omega'$ between the same two configurations, treated now as members of $\mathbf{\Sigma}[S_{13}]$. This path contributes only to $S_{13}$-associated transitions.}]
{\label{subfig:B_13} \includegraphics{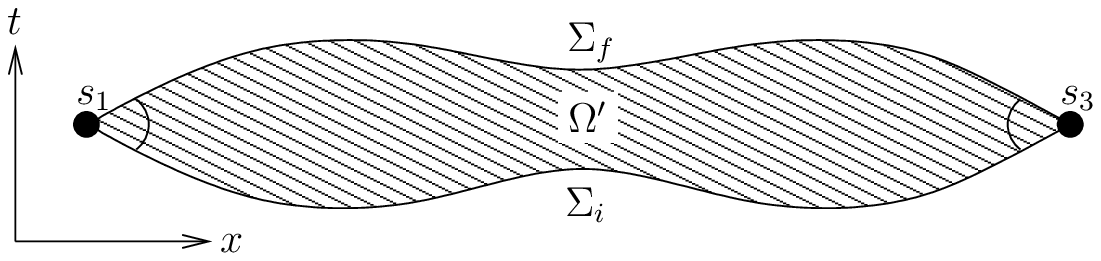}} %
\caption{Cross-section visualizations of the objects used in Sec. \ref{subsec:main:proportionality}.}
\end{figure}

If indeed $\dim{H} < |\mathbf{\Sigma}_C|$, then our last conclusion can be roughly restated as follows:
\begin{reformulation}
The existence of finite paths between any pair of $\Sigma$'s hampers the orthogonality of states, making the mutually orthogonal basis smaller than would be expected in canonical quantization. 
\end{reformulation}
\noindent
This line of reasoning can be extended further, leading to: 
\begin{conjecture} \nonumber
If we increase the number of allowed paths between pairs of $\Sigma$'s, then $\dim{H}$ will become smaller still.
\end{conjecture}
\noindent
Using this idea, we propose a way to determine how $\dim{H[S]}$ may depend on the properties of $S$. Specifically, we compare the two interesting special cases:
\begin{enumerate}
\item Holographic area-proportionality: $\dim{H[S]} \propto A[S]$
\item Canonical volume-proportionality: $\dim{H[S]} \propto V[S]$. Incidentally, with $S$ as the basic variable, this $V$ isn't trivial to define. It will require something like $V[S] = \alpha[S] A[S]^{(d-1)/(d-2)}$, with $\alpha[S]$ a dimensionless shape-dependent coefficient. See Sec. \ref{subsec:reqularization:configuration_space_size} for some discussion on this.
\end{enumerate}
To discriminate between the two possibilities, we imagine a situation in which the area $A[S]$ changes, while the enclosed volume does not. Consider a surface $S_{123} = S_{12} \cup S_{23}$, composed of two compact surfaces with an overlap $s_2$ (see fig. \ref{subfig:S_123}). Then we have:
\begin{eqnarray}
A[S_{123}] &=& A[S_{12}] + A[S_{23}] - A[s_2] \\
V[S_{123}] &=& V[S_{12}] + V[S_{23}] \\
\mathbf{\Sigma}[S_{123}] &=& \mathbf{\Sigma}[S_{12}] \times \mathbf{\Sigma}[S_{23}] \\
H[S_{123}] &=& H[S_{12}] \oplus H[S_{23}]
\end{eqnarray}
Now suppose we remove the overlap, yielding a new surface $S_{13} = S_{12} \cup S_{23} \setminus s_2$ (fig. \ref{subfig:S_13}). Then $V[S_{13}] = V[S_{123}]$, but $A[S_{13}] < A[S_{123}]$. Also, there are now additional paths satisfying a given boundary condition $B$; this is because $\Sigma_i$ and $\Sigma_f$ are no longer constrained to intersect at $s_2$ (figs. \ref{subfig:B_123}, \ref{subfig:B_13}). This alters the values of transition amplitudes, even when $B$ is the same before and after the removal. Specifically, our conjecture indicates that the increased number of allowed paths leads to fewer orthogonalities, resulting in $\dim{H[S_{13}]} < \dim{H[S_{123}]}$. Such behaviour is expected for $H[S] \propto A[S]$, but not for $H[S] \propto V[S]$. This doesn't imply actual area-proportionality, which may not be achieved before the Planck scale. Instead, we have a deviation from volume-proportionality in the desirable direction. To summarize:
\begin{statement}
A reduction of the area $A[S]$ which doesn't affect the enclosed volume might lead to a smaller Hilbert-space dimension $\dim{H[S]}$. This shows that the Hilbert space dimension may begin to exhibit features of area-proportionality already at low energy.
\end{statement}
\noindent

\section{Comparison with ordinary QFT} \label{sec:QFT}

We'll now carry out an analogous discussion in the context of fixed-geometry QFT. In order to emphasize the unique features of the gravitational case, we'll consider a theory such as Yang-Mills, with gauge (but not diffeomorphism) invariance. The theory's fields are decomposed into $\phi = (g, \varphi)$, where $g$ are the static fields defining the geometry, and $\varphi$ are the dynamic fields encoding the degrees of freedom. The physics is played out on a given configuration of the $g$-fields, which is our spacetime $\mathcal{M}$.

Once again, we begin with a codimension-2 surface $S \subset \mathcal{M}$. This time its shape is defined not intrinsically, but through its position in $\mathcal{M}$. There are two null hypersurfaces $L^-_\mathcal{M}[S],\, L^+_\mathcal{M}[S] \subset \mathcal{M}$ bounded by $S$, one past-converging and the other future-converging. In the special case where $\mathcal{M}$ is Minkowski space and $S$ is a metric sphere, $L^\pm_\mathcal{M}[S]$ are lightcones; generally, they terminate at the far end on a codimension-3 caustic. $L^-_\mathcal{M}[S]\, \cup\, L^+_\mathcal{M}[S]$ is a compact hypersurface; it bounds a spacetime region, which we denote by $M[S]$. In other words, \emph{the Lorentzian causal structure maps a $(d-2)$-dimensional surface to a $d$-dimensional spacetime region}. This construction applies to the gravitational case as well, with the difference that the causal structure is state-dependent. Again we see the potential for $S$-dependence of the restricted Hilbert space, from a slightly different perspective than in Sec. \ref{subsec:main:states_transitions}. 

\begin{figure}%
\centering%
\label{fig:M} \includegraphics{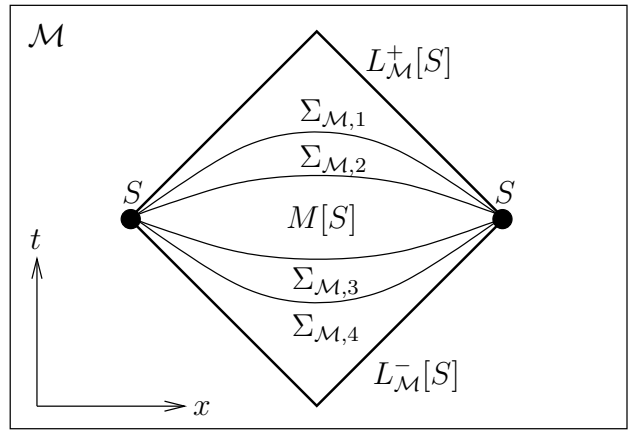}
\caption{The spacetime region $M[S]$ causally defined by the surface $S$ in a given spacetime $\mathcal{M}$. Examples are shown of spacelike hypersurfaces in $\mathbf{\Sigma}_\mathcal{M}[S]$, along with the limiting null hypersurfaces $L^\pm_\mathcal{M}[S]$.}
\end{figure}%
Proceeding with our analogy, let $\mathbf{\Sigma}_\mathcal{M}[S]$ be the set of spacelike hypersurfaces $\Sigma_\mathcal{M} \subset M[S]$ such that $\partial\Sigma_\mathcal{M} = S$. These are the Cauchy hypersurfaces for $M[S]$. $L^\pm_\mathcal{M}[S]$ are the limiting cases of such hypersurfaces; see fig. \ref{fig:M}. On each $\Sigma_\mathcal{M} \in \mathbf{\Sigma}_\mathcal{M}$, the possible field values $\varphi^{(d-1)}(x^i)$ for $x^i \in \Sigma_\mathcal{M}$ constitute a configuration space $\mathbf{\sigma}[\Sigma_\mathcal{M}]$. The configurations in $\mathbf{\sigma}[\Sigma_\mathcal{M}]$ are defined up to $(d-1)$-dimensional gauge transformations, just as the $\Sigma$'s of Sec. \ref{subsec:main:states_transitions} were defined up to $(d-1)$-dimensional gauge transformations and diffeomorphisms. One important difference is that $\mathbf{\sigma}[\Sigma_\mathcal{M}]$ contains the theory's true number of degrees of freedom, while the gravitational $\mathbf{\Sigma}[S]$ contains one too many, requiring a Hamiltonian constraint. In an ordinary gauge theory, all ambiguities in physical states are removed by factoring out the gauge transformations within $\Sigma_\mathcal{M}$. In gravity, an additional gauge freedom remains, regarding the position of $\Sigma$ itself within the spacetime; it is this freedom that the Hamiltonian constraint expresses.

Moving on to quantum mechanics, each configuration space $\mathbf{\sigma}[\Sigma_\mathcal{M}]$ spans a Hilbert space $H[\Sigma_\mathcal{M}]$. Between two configurations $\sigma_i \in \mathbf{\sigma}[\Sigma_{\mathcal{M},i}]$ and $\sigma_f \in \mathbf{\sigma}[\Sigma_{\mathcal{M},f}]$, a transition $\langle\sigma_f;\Sigma_{\mathcal{M},f}|\sigma_i;\Sigma_{\mathcal{M},i}\rangle$ can be defined \footnote{ Since we know the positions of $\Sigma_{\mathcal{M},i}$ and $\Sigma_{\mathcal{M},f}$ within $\mathcal{M}$, there is no need for joining parameters $j(x^a)$.}. These transitions allow us to express states in $H[\Sigma_{\mathcal{M},f}]$ in terms of states in $H[\Sigma_{\mathcal{M},i}]$. This trivial fact of unitary evolution means that all the Hilbert spaces $H[\Sigma_\mathcal{M}]$ for $\Sigma_\mathcal{M} \in \mathbf{\Sigma}_\mathcal{M}[S]$ are in fact the same Hilbert space, which we can denote by $H_\mathcal{M}[S]$. Curiously, we see that in a limited sense, already in ordinary QFT the space of states is defined by a codimension-2 bounding surface. But note the crucial difference from the gravitational case: the Hilbert space is defined here not just by intrinsic properties of $S$, but also by the geometry of its interior. Though the full geometry of $M[S]$ isn't necessary, we do need the geometry of at least one hypersurface $\Sigma_\mathcal{M}$ to construct a copy $H[\Sigma_\mathcal{M}]$ of the Hilbert space. Thus, in contrast to the conclusion of Sec. \ref{subsec:main:states_transitions}, we find that $H$ is a functional of $(d-1)$-dimensional data, rather than the $(d-2)$-dimensional data provided by $S$.

The argument of Sec. \ref{subsec:main:canonical} clearly doesn't apply in a fixed-geometry setup. For each $\Sigma_\mathcal{M} \in \mathbf{\Sigma}_\mathcal{M}[S]$, the configuration space $\mathbf{\sigma}[\Sigma_\mathcal{M}]$ \emph{does} constitute an orthogonal basis for $H_\mathcal{M}[S]$, with an inner product that is a well-defined limit of transition amplitudes. Indeed, consider a transition $\langle\sigma_f;\Sigma_{\mathcal{M},f}|\sigma_i;\Sigma_{\mathcal{M},i}\rangle$. The relative position of $\Sigma_{\mathcal{M},i}$ and $\Sigma_{\mathcal{M},f}$ is entirely fixed; as $\Sigma_{\mathcal{M},i} \rightarrow \Sigma_{\mathcal{M},f}$, the path integral for $\sigma_f \neq \sigma_i$ involves large field derivatives in a region of vanishing size, leading to: 
\begin{eqnarray}
\langle\sigma_f;\Sigma_{\mathcal{M}}|\sigma_i;\Sigma_{\mathcal{M}}\rangle = \delta(\sigma_f - \sigma_i)
\end{eqnarray}
Which yields the familiar Hilbert-space dimension:
\begin{eqnarray}
\dim{H_\mathcal{M}[S]} = N_\varphi(\Lambda)^{\Lambda^{d-1}V[S]}
\end{eqnarray}
In other words, we see that the 'a-priori' and 'dynamical' definitions of the Hilbert-space product coincide.

The construction of Sec. \ref{subsec:main:proportionality} doesn't lead to non-trivial results either. Indeed, consider two hypersurfaces $\Sigma_{\mathcal{M},i}, \Sigma_{\mathcal{M},f} \in \mathbf{\Sigma}_\mathcal{M}[S_{123}]$; they intersect at $S_{123}$, and in particular at $s_2 \subset S_{123}$. Now, the position of $\Sigma_{\mathcal{M},i}, \Sigma_{\mathcal{M},f}$ is fixed by their coordinates in $\mathcal{M}$. As a result, the removal of $s_2$ from $S_{123}$ won't change the fact that $s_2 \subset \Sigma_{\mathcal{M},i} \cap \Sigma_{\mathcal{M},f}$. Therefore, the allowed paths between two configurations on $\Sigma_{\mathcal{M},i}, \Sigma_{\mathcal{M},f}$ are the same whether we treat them as elements of $\mathbf{\Sigma}_\mathcal{M}[S_{123}]$ or of $\mathbf{\Sigma}_\mathcal{M}[S_{13}]$. As desired, the arguments for a holographic $\dim{H}$ apply only to a diffeomorphism-invariant theory.

\section{Comparison with other methods in quantum gravity} \label{sec:methods}

\subsection{Global vs. local path integrals} \label{subsec:methods:global_local}

In QFT, we've become accustomed to using 'global' path integrals of the form:
\begin{eqnarray} \label{eq:global}
\langle0|\mathcal{O}|0\rangle = 
\int{\frac{\mathcal{D}\varphi^{(d)}(x^\mu)}{\mbox{Gauge}[\varphi^{(d)}(x^\mu)]}\; \mathcal{O}[\varphi^{(d)}]\; e^{iS[\varphi^{(d)}]}}
\end{eqnarray}
taken with vacuum boundary conditions at asymptotic infinity. Indeed, in a fixed-geometry theory any amplitude can be expressed in this way. For example, to obtain the amplitudes of Sec. \ref{sec:QFT}, we'd need the following operator:
\begin{eqnarray} \label{eq:O}
\mathcal{O}[\sigma_i,\sigma_f] = \int{\mathcal{D}\mbox{Gauge}[\varphi^{(d-1)}(x^i)] \prod_{x^i \in \Sigma_{\mathcal{M},i}}{\delta(\varphi^{(d-1)}(x^i) - \varphi_{\sigma_i}^{(d-1)}(x^i))} \prod_{x^i \in \Sigma_{\mathcal{M},f}}{\delta(\varphi^{(d-1)}(x^i) - \varphi_{\sigma_f}^{(d-1)}(x^i))}}
\end{eqnarray}

Path integrals of the form \eqref{eq:global} are used for perturbative calculations of the low-energy graviton S-matrix in asymptotically flat spacetime. No holographic properties arise in such calculations, as we will discuss in Sec. \ref{subsec:observables:special}. On the other hand, the holography-oriented results of Sec. \ref{sec:main} stem from our usage of 'local' $B$-bounded path integrals, as in \eqref{eq:transition}. A suspicion may arise that our results are merely an artefact of this approach, and would be invalidated by the correct usage of a global path integral. To counter this objection, we argue that our observables $T_S[B]$ \emph{cannot} in fact be expressed by a global path integral without violating diffeomorphism invariance. In our invariant formulation, the boundary conditions $B$ are defined intrinsically. This is no longer possible if we are to use a global path integral: an operator $\mathcal{O}$ is defined in terms of spacetime coordinates $x^\mu$; we'd have to assign such coordinates to $\Sigma_i$ and $\Sigma_f$, i.e. choose an embedding for $B$. From the standpoint of general covariance, this is clearly undesirable. More specifically, our operator $\mathcal{O}[B]$ wouldn't be invariant under the full diffeomorphism group. A definition along the lines of \eqref{eq:O} can be made invariant under transformations which leave the points of $B$ intact, while perhaps permuting them among themselves; but we see no way to make $\mathcal{O}[B]$ invariant under transformations which alter the embedding of $B$; these would cause the operator to evaluate fields in other parts of the spacetime \footnote{Incidentally, diffeomorphisms which alter the embedding of boundary hypersurfaces are also the reason for a Hamiltonian constraint in canonical quantum gravity.}. The exception to this is when $B$ is a natural boundary of the spacetime, as is the case in S-matrix calculations. Now, the known approaches to path integrals in a gauge-invariant theory, such as Fadeev-Popov and BRST, require a gauge-invariant operator in the integrand. As we have seen, this isn't possible when evaluating a finite-region transition amplitude with diffeomorphisms as the gauge group. Apparently, the observables of Sec. \ref{sec:main} require the usage of $B$-bounded path integrals.

\subsection{Canonical quantum gravity} \label{subsec:methods:canonical}

Canonical quantum gravity is the traditional approach to the quantization of General Relativity in terms of field configurations on spacelike hypersurfaces \cite{isham:1992ms}. In CQG, codimension-1 configurations are used as a Hilbert basis, in contradiction to the conclusions of Sec. \ref{subsec:main:canonical}. To settle the matter, we'd like to formulate CQG within the framework of Sec. \ref{sec:main}, and carry out a comparison between the two approaches. We'll find that the difference lies in an unremovable gauge dependence of CQG, which renders it irrelevant if we insist on diffeomorphism invariance.

CQG deals with configurations on spacelike hypersurfaces, much like the $\Sigma$'s of Sec. \ref{sec:main}. It usually addresses all of space rather than a portion of it, which puts the boundary surface $S$ at spatial infinity. To capture the true number of degrees of freedom, the configuration space $\mathbf{\Sigma}$ is reduced to a subspace $\mathbf{\Sigma}_C \subset \mathbf{\Sigma}$ by imposing a codimension-1 Hamiltonian constraint $C$. Additional codimension-2 boundary conditions are imposed on $\mathbf{\Sigma}_C$, usually to make the $\Sigma$'s asymptotically flat and parallel. When considering several $\Sigma$'s, this allows us to assign a time coordinate $t$ to each of them, up to a global shift. Thus $\mathbf{\Sigma}_C$ is reduced to 'fixed-time' subspaces $\mathbf{\Sigma}_C(t)$ \footnote{Technically, each $\mathbf{\Sigma}_C(t)$ is a copy of the \emph{same} subspace of $\mathbf{\Sigma}_C$, defined e.g. by asymptotic flatness.}. The 'time' $t$ is an asymptotic parameter, which describes the hypersurface's boundary rather than its bulk. In the language of Sec. \ref{sec:main}, the time difference between two hypersurfaces $\Sigma_i \in \mathbf{\Sigma}_C(t_i),\; \Sigma_f \in \mathbf{\Sigma}_C(t_f)$ defines a joining function $j^{(d-2)}(x^a; t_f - t_i)$ on the asymptotic boundary $S$. In particular, two $\Sigma$'s with the same $t$ are joined by the 'identity' joining function $j_0$.

CQG goes on to postulate that each $\mathbf{\Sigma}_C(t)$ forms an orthogonal Hilbert basis, an 'a-priori' definition of the Hilbert space. As discussed in Sec. \ref{subsec:main:canonical}, this stands in contradiction to the transition amplitudes calculated with an invariant path integral. The amplitude $\langle\Sigma_f|\Sigma_i\rangle_j$ for general $\Sigma_i,\Sigma_f \in \mathbf{\Sigma}$ and $j \in \mathbf{J}$ consists of the available paths within the boundary $B[\Sigma_i,\Sigma_f,j]$. In particular, in the classical limit it takes the form $\rho\,e^{iS}$, where $S$ is the action of the classical path; $\rho$ is related to the second derivative of the action with respect to the path, which in turn is related to the field derivatives involved. When the classical path evolves two different $\Sigma$'s into each other over a vanishing spacetime volume, $\rho$ tends to zero. Now, in a diffemorphism-invariant theory with no fixed geometry, it's not easy to make the spacetime volume between two $\Sigma$'s vanish. Choosing $\Sigma$'s with the same $t$ value isn't enough: this way only the joining parameters are constrained, and paths with a finite hypervolume are still allowed. The $(d-1)$-dimensional Hamiltonian constraint doesn't change this situation: without a background spacetime, a single equation \emph{within} each $\Sigma$ isn't enough to make them coincide. An exception to this is the limiting case where the constraint selects those hypersurfaces that are \emph{null} rather than spacelike; we'll elaborate on this in Sec. \ref{sec:observables}. Once again, we've seen that a configuration space subject to a Hamiltonian constraint isn't $T_S$-orthogonal, so our consistency requirement $P_S \mapsto T_S$ disqualifies it as a Hilbert basis. 

It may seem strange that the correction brought by $C$ to the number of degrees of freedom counts for so little in the above argument. What happens here is that the degrees-of-freedom count is correct only on average, without an actual one-to-one correspondence. Even if the majority of spacetime configurations contains exactly one member of each $\mathbf{\Sigma}_C(t)$, it's not forbidden for a spacetime to contain more than one, or none at all. This is an aspect of the 'floppiness' brought by diffeomorphism invariance. The non-vanishing paths which make $\Sigma_1,\Sigma_2 \in \mathbf{\Sigma}_C(t)$ non-orthogonal come from that minority of spacetimes which contain both $\Sigma_1$ and $\Sigma_2$, despite the average tendency to have only one $\Sigma \in \mathbf{\Sigma}_C(t)$.

Finally, it may help to draw the comparison between our approach and CQG on the common ground of path integrals. CQG is defined by a Hamiltonian, which evolves $\Sigma \in \mathbf{\Sigma}_C(t)$ into a superposition of basis elements in $\mathbf{\Sigma}_C(t+dt)$. This infinitesimal evolution can be exponentiated, leading to a path-integral formulation. What is the difference, then, between this path integral and ours? Let us restrict ourselves to the observables of CQG, i.e. to transitions only between members of the $\mathbf{\Sigma}_C(t)$ spaces. Then our path integral \eqref{eq:transition} contains all the possible paths between $\Sigma_i \in \mathbf{\Sigma}_C(t_i)$ and $\Sigma_f \in \mathbf{\Sigma}_C(t_f)$. Note that $C$ is imposed here only on the \emph{endpoints} of the path. The CQG path integral, on the other hand, is built from the infinitesimal evolution of $C$-constrained hypersurfaces one into the other. Therefore it contains not all the paths between $\Sigma_i$ and $\Sigma_j$, but only those in which they are connected by a continuous series $\Sigma(t) \in \mathbf{\Sigma}_C(t)$, with $\Sigma_i = \Sigma(t_i)$ and $\Sigma_f = \Sigma(t_f)$. Now, since we deal with smooth geometries, any hypersurface in $\mathbf{\Sigma}_C(t_{i,f})$ is part of such a series. However, recall that a minority of paths contain several members of the same $\mathbf{\Sigma}_C(t)$. For such paths, $\Sigma_i$ and $\Sigma_f$ may not lie in the same continuous series; these paths will contribute to the invariant integral \eqref{eq:transition}, but not to the CQG integral. We see that within the common scope of applicability, CQG and the invariant approach give different transition amplitudes. Furthermore, CGQ's answers depend on the constraint $C$. Indeed, for two different constraints $C$ and $C'$, it's possible to find hypersurfaces $\Sigma_i,\Sigma_f$ which satisfy both of them, and a path for which these $\Sigma_i,\Sigma_f$ are on the same continuous $C$-series but on different $C'$-series. This path will contribute to the $C$-based path integral, but not to the $C'$-based path integral. But the choice of $C$ is nothing but a partial choice of gauge. We see that the amplitudes calculated in CQG are gauge-dependent, and differ from those given by an invariant path integral. The canonical basis is thus unsuitable for a theory with full diffeomorphism invariance, and we should use the 'dynamical' Hilbert-space definition instead.

\section{Comparison with other observables} \label{sec:observables}

In the previous section, we discussed other approaches to calculating the amplitudes of Sec. \ref{sec:main}. We argued that our results are not the artefact of a misplaced method, but the natural result of diffeomorphism invariance and causality when applied to our observables. The next question is how these results depend on the choice of observables. Like most other aspects, this choice differs among various treatments of quantum gravity. Having identified some alternative invariant observables, we will find that the results \emph{do} depend on the amplitudes we choose to calculate; two factors stand out in this respect:
\begin{enumerate}
\item The causal structure of the boundary conditions.
\item Whether the boundary conditions are global in both space and time, or refer to a localized partial region. 
\end{enumerate}

\subsection{Transition amplitudes vs. field correlations}
\label{subsec:observables:correlations}

We've been discussing transition amplitudes with compact codimension-1 boundary conditions. It can be argued that such amplitudes describe all possible experiments in quantum mechanics. However, in the formalism of QFT it's often convenient to express observables in terms of vacuum field correlations; these are calculated with a path integral of the form \eqref{eq:global}, usually with a product of fields for $\mathcal{O}$:
\begin{eqnarray} \label{eq:field_product}
\mathcal{O} = \phi(x_1^\mu)\phi(x_2^\mu)\phi(x_3^\mu)\dots
\end{eqnarray}
Such correlations are used so often that they become treated as observables in their own right. In ordinary QFT, this blurring of concepts is quite harmless. Path integrals of the form \eqref{eq:global} can describe any transition amplitude; in fact they are richer, since they include off-shell amplitudes as well. Furthermore, specific kernel operators such as \eqref{eq:field_product} provide us with simple questions, with the hope for an easily calculable answer.

It is therefore tempting to extend this successful method to quantum gravity, i.e. to take field-product correlations, rather than hypersurface transition amplitudes, as our primary observables. However, this is incompatible with diffeomorphism invariance. The reason is the same as in Sec. \ref{subsec:methods:global_local}. Being a function of fields at points with given coordinates, the operator $\mathcal{O}$ cannot be made diffeomorphism-invariant. Conceivably, this difficulty can be overcome. For instance, we may choose to calculate an amplitude $\langle0|\phi(x_1^\mu)\phi(x_2^\mu)|0\rangle$, where the points $x_1^\mu,x_2^\mu$ are defined not by their coordinates, but by the geodesic distance between them. However, such an observable would still possess another undesirable property, one that is shared by most choices of $\mathcal{O}$. Our amplitude would be assembled from contributions of the possible spacetimes around the two points, and this set of spacetimes ultimately depends on our global boundary conditions. Unlike ordinary QFT, in quantum gravity it's quite unclear what a 'natural' boundary condition should be, and in any case a dependence on it can be viewed as a breach of general covariance. This problem can be avoided if our observable is somehow independent of the surrounding larger spacetime. We conclude that a diffeomorphism-invariant 'question', i.e. the input data for a calculation, should preferably satisfy two criteria:
\begin{enumerate}
\item It should be defined intrinsically, i.e. without reference to an embedding coordinate system.
\item It should delimit a causally isolated region, unaffected by the shape of the surrounding spacetime.
\end{enumerate}
The amplitudes of Sec. \ref{sec:main} satisfy these conditions. Field correlations do not, so we discard them in our discussion of invariant observables. Note, however, the successful use in \cite{Modesto:2005sj,Rovelli:2005yj} of field correlations \emph{in conjunction} with boundary conditions of the type decribed in Sec. \ref{sec:main}.

\subsection{The S-matrix and other special transitions} \label{subsec:observables:special}

The generic transitions of Sec. \ref{sec:main} can be specialized in two directions: by using special $\Sigma$'s given a generic $S$, or by using a special $S$ to begin with. In an asymptotically-flat spacetime, the spacelike infinity provides a natural choice for such a special $S$. With this $S$ we can discuss the evolution of spacelike hypersurfaces spanning the entire space. As discussed in Sec. \ref{subsec:methods:canonical}, this is easily incorporated as a special case of our general approach, without affecting the conclusions of Sec. \ref{sec:main}. However, if we considered \emph{only} such global hypersurfaces, we would lose the $S$-dependence insights of Secs. \ref{subsec:main:states_transitions} and \ref{subsec:main:proportionality} (the results of Sec. \ref{subsec:main:canonical} would remain intact). Furthermore, as part of our expectations from a physical theory, we believe it should describe not just global processes, but also processes in limited regions of space. As argued in Sec. \ref{subsec:methods:global_local}, for quantum gravity this cannot be achieved with global path integrals. Therefore, we should insist on transition amplitudes with generic finite $S$, leaving global states as a consistent limiting case.

Now let's turn our attention to special transitions associated with a given $S$. As discussed in Sec. \ref{sec:QFT}, in a given geometry $\mathcal{M}$, each $S$ is associated with two null hypersurfaces $L^\pm_\mathcal{M}[S]$; these are the outer limiting cases of spacelike hypersurfaces $\Sigma_\mathcal{M} \in \mathbf{\Sigma}_\mathcal{M}[S]$. Such null hypersurfaces can be used in a diffeomorphism-invariant theory as well. Let $\mathbf{L}^-[S]$ be the space of past-converging null-hypersurface configurations bounded by $S$, and let $\mathbf{L}^+[S]$ be the analogous space for future-converging ones. These configuration spaces possess the theory's true number of degrees of freedom, much like the constrained spaces $\mathbf{\Sigma}_C$ from Sec. \ref{subsec:methods:canonical}. But unlike the $\mathbf{\Sigma}_C$'s, the null spaces $\mathbf{L}^\pm[S]$ really stand in a one-to-one correspondence with the theory's states \cite{penrose:1988}: while a spacelike $\Sigma \in \mathbf{\Sigma}_C$ may appear any number of times in a particular path, there's always just one causal boundary $L^- \in \mathbf{L}^-[S]$ in its past and one $L^+ \in \mathbf{L}^+[S]$ in its future. Equivalently, there are no paths of finite hypervolume containing two different hypersurfaces $L_1^-,L_2^- \in \mathbf{L}^-[S]$ (and likewise for $\mathbf{L}^+[S]$). 

Suppose now that the transition amplitudes within and between $\mathbf{L}^\pm[S]$ are taken to be our only observables. Let's examine the impact of this on the results of Sec. \ref{sec:main}. From the previous paragraph, we see that the argument of Sec. \ref{subsec:main:canonical} no longer applies: both $\mathbf{L}^-[S]$ and $\mathbf{L}^+[S]$ constitute an orthogonal Hilbert basis, spanning a Hilbert space $H_L[S]$ of the conventional dimensionality $N_\phi(\Lambda)^{\Lambda^{d-1}V[S]}$. The unitary transformation between these two bases is given by the transition amplitudes $\langle L^+|L^- \rangle$. The argument of Sec. \ref{subsec:main:states_transitions} is left intact: the configuration spaces $\mathbf{L}^+$, and therefore the Hilbert space $H_L$, are still functionally determined by $S$; however, this does nothing to change the non-holographic value of $\dim{H_L}$. Finally, if only null hypersurfaces are considered, the argument of Sec. \ref{subsec:main:proportionality} becomes irrelevant: $\mathbf{L}^\pm[S_{13}]$ has nothing to do with $\mathbf{L}^\pm[S_{12}] \times \mathbf{L}^\pm[S_{23}]$, so no simple comparison can be drawn between $H_L[S_{123}]$ and $H_L[S_{13}]$. 

As an example, consider pure quantum gravity on asymptotically-flat spacetime, with $S$ again taken at spatial infinity. Then $\mathbf{L}^\pm[S]$ refer to the past and future null infinities $\mathcal{I}^\pm$. The Hilbert space $H_L$ spanned by $\mathbf{L}^\pm[S]$ is the familiar Fock space of gravitons. Transition amplitudes of the form $\langle L^+|L^- \rangle$ constitute the S-matrix. In agreement with this section's reasoning, no holographic behavior is known in this relatively well-studied system of observables \cite{Burgess:2003jk,Nakanishi:1982ag}. 

At first sight, therefore, null-hypersurface transitions invalidate altogether the results of Sec. \ref{subsec:main:canonical}. We claimed there that $\dim{H[S]}$ \emph{might} be smaller than its canonical value, and that we don't understand enough to say for sure; but given a well-understood special case in which the canonical dimension \emph{does} work, the argument seems lost. Indeed, the $\Sigma$-states can be expressed in terms of the $\mathbf{L}^-[S]$ basis through the transition amplitudes $\langle\Sigma|L^-\rangle$ (and similarly for the $\mathbf{L}^+[S]$ basis). This seems to place all the $\Sigma$'s within the Hilbert space $H_L[S]$, whose dimension is non-holographic. Yet there is a subtlety here, which reaffirms the possibility of a smaller $\dim{H}$. In our approach, the primary content of the theory are the transition amplitudes $T_S:\; \langle\Sigma_f|\Sigma_i\rangle_j$, while the Hilbert-space structure $H[S]$ is required to emerge from them. Now, while the $\Sigma$'s can indeed be expressed in terms of $H_L[S]$, their transition amplitudes cannot. The reason is that amplitudes such as $\langle L^+|L^-\rangle$, $\langle\Sigma|L^-\rangle$ and $\langle L^+|\Sigma\rangle$ don't require a joining function $j$: due to the special status of $L^\pm$'s as causal boundaries, there's no need for joining parameters as a separate input. For example, if $j(x^a)$ is the angle between $\Sigma_i, \Sigma_f$ at $x^a \in S$, then for null hypersurfaces $j(x^a) = \infty$ always, i.e. it doesn't add new information to the amplitude's boundary conditions. As a result, amplitudes of the form $\langle\Sigma|L^-\rangle$ and $\langle L^+|\Sigma\rangle$ cannot be used to calculate the generic amplitudes $\langle\Sigma_f|\Sigma_i\rangle_j$: the structure of null-hypersurface transitions isn't rich enough to distinguish between different $j$'s. In other words, the generic amplitudes $T_S$ cannot be derived from $H_L[S]$. These amplitudes must therefore be defined on their own terms (as in Sec. \ref{sec:main}); a Hilbert-space structure $H[S]$ is then derived from them, independently of $H_L[S]$.

We found that the Hilbert-space structure $H[S]$ of spacelike-hypersurface transitions can't be derived directly from the null-hypersurface Hilbert space $H_L[S]$. Nonetheless, null hypersurfaces are a limit of spacelike ones; naively, this suggests that $H[S]$ and $H_L[S]$ must still be equivalent. However, the situation here is more involved. To obtain $H_L[S]$ as a limit of generic amplitudes $T_S$, two \emph{separate} limits must be taken: 
\begin{itemize}
\item The boundary hypersurfaces should become null: $\Sigma \rightarrow L^\pm$
\item The joining function should approach the identity: $j \rightarrow j_0$; this condition produces the mapping $P_S \mapsto T_S$ between the Hilbert-space product and the transition amplitudes. 
\end{itemize}
These two limits may be contradictory. For example, consider again $j$ as a joining angle. For null hypersurfaces, angles with other hypersurfaces become infinite. Thus the first condition implies $j \rightarrow \infty$, while the second implies $j \rightarrow 0$. Therefore, the order in which we take these limits is important. The non-holographic $H_L[S]$ from the previous paragraphs results from taking $j \rightarrow j_0$ before $\Sigma \rightarrow L^\pm$. Perhaps this is a distortion of the $T_S$-based $H[S]$, resulting from the limit's singularity. If so, the correct order may be to take $\Sigma \rightarrow L^\pm$ before $j \rightarrow j_0$, which would correctly place the null-hypersurface states within $H[S]$; then the holography-oriented arguments of Sec. \ref{sec:main} are back on the table. In summary, choosing null-hypersurface transitions as the primary observable leads to a non-holographic Hilbert space, but this is not necessarily the case if they are treated as a limit of the more general spacelike transitions. 

As with the question of global vs. generic $S$, we believe that transitions involving spacelike $\Sigma$'s are a necessary ingredient of the theory. Our physics ought to describe continuous evolution; S-matrix-like 'end-to-end' amplitudes such as $\langle L^+|L^- \rangle$ cannot achieve this. We should therefore allow generic-boundary transitions, like those defined in Sec. \ref{sec:main}. Null-boundary transitions should be treated as their consistent limit, rather than a self-contained set of observables.

\subsection{Spacetimes with different causal structure} \label{subsec:observables:different_causal}

In the previous subsection, we've witnessed the importance of the boundary's causal structure for our arguments. But so far we've considered only those causal structures which correspond to regions of nearly-flat spacetime. Other boundaries can be examined, such as the well-studied AdS. There we have a timelike asymptotic hypersurface at spatial infinity. Since this boundary is timelike, it can serve as a stage for dynamics, as it does in the AdS/CFT correspondence. There is no analogue for this in the constructions of Sec. \ref{sec:main}. Note that unlike our transitions which can be defined on partial regions of space, the AdS boundary can work only if it is the boundary of the entire spacetime. This is because a timelike-bounded spacetime region would be causally influenced from outside, unless there \emph{is} no outside.

Some work has also been done on dS-shaped boundaries. Again, no analogue for these exists in our constructions. A comparison has been made between the behaviour of quantum gravity within AdS, dS and Minkowski global boundaries, and each of them gives quite a different picture of the theory's observables \cite{Witten:2001kn}. In this paper, we explored the consequences of a boundary structure different from these three, which seems more natural for finite regions and small curvature. Incidentally, of the three global boundaries cited above, the global Minkowski boundary with its S-matrix is the only one for which there's no evidence for holography at all. In Sec. \ref{subsec:observables:special}, we've expressed this case as a limit of our more general setup, and argued that its non-holographic Hilbert space may be the result of taking this limit incorectly.

\section{A note on diffeomorphism-invariant low-energy regularization} \label{sec:regularization}

\subsection{Regularization} \label{subsec:reqularization:reqularization}

In the above, we've been assuming a diffeomorphism-invariant, low-energy regularization at an inverse length scale $\Lambda$. A short discussion on the meaning and possible realizations of this assumption is in order. 

Regularization for general boundary conditions on finite hypersurfaces is analytically difficult already in ordinary QFT. That is because such boundaries do not respect symmetries such as translation invariance. The problem is usually approached with numerical reqularizations of the Wilson type, which involve the discretization of spacetime. Lattice QCD is an important successful application of these ideas. In a diffeomorphism-invariant theory as well, discretizations of spacetime provide the most plausible path towards the regularization we're looking for. 

A conceptual problem arises: what is the meaning of a cutoff scale when there is no pre-given metric? In other words, how can we give meaning to the lattice spacing? To begin to answer this, note that we are never really left without a scale: we do know the area, and perhaps other properties, of $S$. This gives us the first meaning of $\Lambda$: it is the resolution at which the boundary data at $S$ are given. But the cutoff scale must also be applied to the hypersurfaces $\Sigma$ and to the path integral. On the conceptual level and staying within the continuous field approach, the cutoff can be formulated as an a-posteriori filtering rule:
\begin{enumerate}
\item Choose a $(d-1)$-dimensional field configuration for $\Sigma$ (or a $d$-dimensional configuration fot the path integral).
\item Examine the configuration in terms of its own metric (or whatever field is responsible for fixing scales). If the fields fluctuate significantly over distances shorter than $1/\Lambda$, discard the configuration. Otherwise, include it in the configuration space (or path integral).
\end{enumerate}
For a more constructive approach, we probably need to discretize spacetime. Promising directions are given by Regge calculus, which is successful in classical gravity \cite{Gentle:2002ux}, and by spin foam models; see \cite{Bonzom:2009hw} and its references. See \cite{Ambjorn:1996ny} for a mathematical discussion on triangulations as approximations of Riemannian manifolds within the context of the Dynamical Triangulations approach. Note that we are interested here in a coarse-grained discretization, rather than the fundamental, quantum discretization arising in the loop quantum gravity, spin foam and dynamical triangulation programs. See \cite{Henson:2009fy} and its references on the transition from such fundamental discretizations to coarse-grained ones. 

At the core of the diffeomorphism-invariant discretizations stands the idea that they start out as combinatorial, i.e. as sets of points defined only by their connectivity to each other and to the boundary. This mimics the differentiable structure of the manifold, which is the only input in the bulk. The distances between the points are then determined by the fields' values on them. The physical spacing between the points can thus be wildly inhomogeneous, but such possibilities may be dynamically suppressed.

\subsection{Size of the configuration space} \label{subsec:reqularization:configuration_space_size}

We have assumed in Sec. \ref{sec:main} that the size $|\mathbf{\Sigma}[S]|$ of the configuration space scales in the naive, non-holographic manner. We've formulated a regularized version of this statement as eq. \eqref{eq:configuration_space_size}. We now briefly address its validity, and the ambiguity involved.

In all the approaches to regularization discussed in Sec. \ref{subsec:reqularization:reqularization}, we need $\sim \Lambda^{d-1}V$ sets of data to describe a hypersurface of volume $V$ to resolution $\Lambda$. This agrees with the naive perturbation theory around a 'background' hypersurface, and with the intuition of \eqref{eq:configuration_space_size}. But what should we take as a generic value for $V$? For spacelike hypersurfaces, an order-of-magnitude approximation may be the volume of the hypersurface $\Sigma_0[S]$ on which a $(d-1)$-dimensional restriction of the action is extremal (assuming enough boundary data on $S$); for pure gravity, this means the Ricci-flat hypersurface. For null hypersurfaces, whose metric volume is zero, this idea breaks down. The problem here is that the coordinate normal to $S$ (let's call it $r$) goes along the lightrays, and has zero length. This also raises the question of how to define the cutoff scale along that coordinate. A possible answer is to attribute length to the normal coordinate not directly from the metric, but by the decrease in the area $A[S]$ as $S$ is deformed into a nearby surface $S'$ by moving it along $r$. This would lead to a characteristic volume $V[S] \sim A[S]^{(d-1)/(d-2)}$ for spacelike as well as null hypersurfaces.

In summary, eq. \eqref{eq:configuration_space_size} for the size of the configuration makes sense on a rudimentary level, but there is serious work to be done before it can be made more precise. 

\section{Prospects for quantitative results} \label{sec:quantitative}

The discussion above has been quite abstract and speculative. A proper test of its validity at a quantitative level presents major difficulties, outlined in Sec. \ref{sec:regularization}. 

However, in the ending of Sec. \ref{subsec:observables:special}, we may have a possible route to quantitative results. There, we consider the disrepancy between our general holography-oriented claims and the known results for the asymptotic graviton Fock space at null infinity. We then propose to blame this discrepancy on an incorrect order of limits. Now, in this case the boundary conditions are asymptotic and thus simple, of the kind we know how to handle analytically. One can attempt to calculate amplitudes of the form $\langle L_2^-|L_1^-\rangle$ with the limits taken in reversed order (spacelike hypersurfaces become null, \emph{and then} they come together). These amplitudes would dynamically define the Hilbert space, and allow a test for our guesses of a smaller-than-canonical Hilbert-space dimension.

\section{Acknowledgements}

I would like to thank Yaron Oz for valuable discussions. This work is supported in part by GIF, by the Israeli Science Foundation center of excellence, by the Deutsch-Israelische Projektkooperation and by the US-Israel binational science foundation.

\bibliographystyle{JHEP}
\bibliography{paper}

\providecommand{\href}[2]{#2}\begingroup\raggedright\begin{thebibliography}{10}

\bibitem{Bekenstein:2000ai}
J.~D. Bekenstein, {\it {Holographic bound from second law of thermodynamics}},
  {\em Phys. Lett.} {\bf B481} (2000) 339--345,
  [\href{http://xxx.lanl.gov/abs/hep-th/0003058}{{\tt hep-th/0003058}}].

\bibitem{Susskind:1994vu}
L.~Susskind, {\it {The World as a hologram}},  {\em J. Math. Phys.} {\bf 36}
  (1995) 6377--6396, [\href{http://xxx.lanl.gov/abs/hep-th/9409089}{{\tt
  hep-th/9409089}}].

\bibitem{'tHooft:1999bw}
G.~'t~Hooft, {\it {The holographic principle: Opening lecture}},
  \href{http://xxx.lanl.gov/abs/hep-th/0003004}{{\tt hep-th/0003004}}.

\bibitem{Bousso:1999dw}
R.~Bousso, {\it {The holographic principle for general backgrounds}},  {\em
  Class. Quant. Grav.} {\bf 17} (2000) 997--1005,
  [\href{http://xxx.lanl.gov/abs/hep-th/9911002}{{\tt hep-th/9911002}}].

\bibitem{Brustein:1999md}
R.~Brustein and G.~Veneziano, {\it {A Causal Entropy Bound}},  {\em Phys. Rev.
  Lett.} {\bf 84} (2000) 5695--5698,
  [\href{http://xxx.lanl.gov/abs/hep-th/9912055}{{\tt hep-th/9912055}}].

\bibitem{Flanagan:1999jp}
E.~E. Flanagan, D.~Marolf, and R.~M. Wald, {\it {Proof of Classical Versions of
  the Bousso Entropy Bound and of the Generalized Second Law}},  {\em Phys.
  Rev.} {\bf D62} (2000) 084035,
  [\href{http://xxx.lanl.gov/abs/hep-th/9908070}{{\tt hep-th/9908070}}].

\bibitem{Aharony:1999ti}
O.~Aharony, S.~S. Gubser, J.~M. Maldacena, H.~Ooguri, and Y.~Oz, {\it {Large N
  field theories, string theory and gravity}},  {\em Phys. Rept.} {\bf 323}
  (2000) 183--386, [\href{http://xxx.lanl.gov/abs/hep-th/9905111}{{\tt
  hep-th/9905111}}].

\bibitem{Rovelli:1996ti}
C.~Rovelli, {\it {Loop quantum gravity and black hole physics}},  {\em Helv.
  Phys. Acta} {\bf 69} (1996) 582--611,
  [\href{http://xxx.lanl.gov/abs/gr-qc/9608032}{{\tt gr-qc/9608032}}].

\bibitem{Yurtsever:2003ii}
U.~Yurtsever, {\it {The holographic entropy bound and local quantum field
  theory}},  {\em Phys. Rev. Lett.} {\bf 91} (2003) 041302,
  [\href{http://xxx.lanl.gov/abs/gr-qc/0303023}{{\tt gr-qc/0303023}}].

\bibitem{Iyer:1994ys}
V.~Iyer and R.~M. Wald, {\it {Some properties of Noether charge and a proposal
  for dynamical black hole entropy}},  {\em Phys. Rev.} {\bf D50} (1994)
  846--864, [\href{http://xxx.lanl.gov/abs/gr-qc/9403028}{{\tt
  gr-qc/9403028}}].

\bibitem{Wald:1993nt}
R.~M. Wald, {\it {Black hole entropy is the Noether charge}},  {\em Phys. Rev.}
  {\bf D48} (1993) 3427--3431,
  [\href{http://xxx.lanl.gov/abs/gr-qc/9307038}{{\tt gr-qc/9307038}}].

\bibitem{Oeckl:2003vu}
R.~Oeckl, {\it {A 'general boundary' formulation for quantum mechanics and
  quantum gravity}},  {\em Phys. Lett.} {\bf B575} (2003) 318--324,
  [\href{http://xxx.lanl.gov/abs/hep-th/0306025}{{\tt hep-th/0306025}}].

\bibitem{Oeckl:2003pw}
R.~Oeckl, {\it {Schroedinger's cat and the clock: Lessons for quantum
  gravity}},  {\em Class. Quant. Grav.} {\bf 20} (2003) 5371--5380,
  [\href{http://xxx.lanl.gov/abs/gr-qc/0306007}{{\tt gr-qc/0306007}}].

\bibitem{Pfeiffer:2003tx}
H.~Pfeiffer, {\it {Diffeomorphisms from finite triangulations and absence of
  'local' degrees of freedom}},  {\em Phys. Lett.} {\bf B591} (2004) 197--201,
  [\href{http://xxx.lanl.gov/abs/gr-qc/0312060}{{\tt gr-qc/0312060}}].

\bibitem{Modesto:2005sj}
L.~Modesto and C.~Rovelli, {\it {Particle scattering in loop quantum gravity}},
   {\em Phys. Rev. Lett.} {\bf 95} (2005) 191301,
  [\href{http://xxx.lanl.gov/abs/gr-qc/0502036}{{\tt gr-qc/0502036}}].

\bibitem{Rovelli:2005yj}
C.~Rovelli, {\it {Graviton propagator from background-independent quantum
  gravity}},  {\em Phys. Rev. Lett.} {\bf 97} (2006) 151301,
  [\href{http://xxx.lanl.gov/abs/gr-qc/0508124}{{\tt gr-qc/0508124}}].

\bibitem{isham:1992ms}
C.~J. Isham, {\it {Canonical quantum gravity and the problem of time}},
  \href{http://xxx.lanl.gov/abs/gr-qc/9210011}{{\tt gr-qc/9210011}}.

\bibitem{penrose:1988}
R.~Penrose and W.~Rindler, {\em Spinors and Space-Time: Volume 1, Two-Spinor
  Calculus and Relativistic Fields (Cambridge Monographs on Mathematical
  Physics)}, ch.~5.11.
\newblock {Cambridge University Press}, February, 1987.

\bibitem{Burgess:2003jk}
C.~P. Burgess, {\it {Quantum gravity in everyday life: General relativity as an
  effective field theory}},  {\em Living Rev. Rel.} {\bf 7} (2004) 5,
  [\href{http://xxx.lanl.gov/abs/gr-qc/0311082}{{\tt gr-qc/0311082}}].

\bibitem{Nakanishi:1982ag}
N.~Nakanishi, {\it {Manifestly covariant canonical formalism of quantum
  gravity: systematic presentation of the theory}},  {\em Publ. Res. Inst.
  Math. Sci. Kyoto} {\bf 19} (1983) 1095.

\bibitem{Witten:2001kn}
E.~Witten, {\it {Quantum gravity in de Sitter space}},
  \href{http://xxx.lanl.gov/abs/hep-th/0106109}{{\tt hep-th/0106109}}.

\bibitem{Gentle:2002ux}
A.~P. Gentle, {\it {Regge calculus: a unique tool for numerical relativity}},
  {\em Gen. Rel. Grav.} {\bf 34} (2002) 1701--1718,
  [\href{http://xxx.lanl.gov/abs/gr-qc/0408006}{{\tt gr-qc/0408006}}].

\bibitem{Bonzom:2009hw}
V.~Bonzom, {\it {Spin foam models for quantum gravity from lattice path
  integrals}},  \href{http://xxx.lanl.gov/abs/0905.1501}{{\tt
  arXiv:0905.1501}}.

\bibitem{Ambjorn:1996ny}
J.~Ambjorn, M.~Carfora, and A.~Marzuoli, {\it {The geometry of dynamical
  triangulations}},  {\em Lect. Notes Phys.} {\bf 50} (1997) 197,
  [\href{http://xxx.lanl.gov/abs/hep-th/9612069}{{\tt hep-th/9612069}}].

\bibitem{Henson:2009fy}
J.~Henson, {\it {Coarse graining dynamical triangulations: a new scheme}},
  \href{http://xxx.lanl.gov/abs/0907.5602}{{\tt arXiv:0907.5602}}.

\end{thebibliography}\endgroup

\end{document}